\documentclass[epjCONF]{svjour}
\usepackage{graphics}
\usepackage[varg]{txfonts} 
\usepackage[latin1]{inputenc}
\usepackage{subfigure}
\session-title{International Conference on New Frontiers in Physics}
\begin{document}
\title{Open heavy-flavour measurements in pp and Pb-Pb collisions with ALICE at the LHC}
\author{Sarah LaPointe\fnmsep\thanks{\email{s.lapointe@cern.ch}} on behalf of the ALICE collaboration}
\institute{ERC- Research Group QGP-ALICE, Utrecht University, Princetonplein 5,
3584 CC Utrecht, The Netherlands}
\abstract{
We present an overview of measurements related to open heavy-flavour production with the ALICE experiment at the LHC. Studies are performed using single leptons (electrons at mid-rapidity and muons at forward-rapidity) and D mesons, which are reconstructed via their hadronic decay channels. The measured differential production cross sections in proton-proton collisions  at $\sqrt{s}$ = 2.76 and 7 TeV are in agreement with perturbative QCD calculations. Results from Pb-Pb collisions at $\sqrt{s_{\mathrm{NN}}}$ = 2.76 TeV on the nuclear modification factor $R_{\mathrm{AA}}$ are shown, along with the elliptic flow $\nu_2$.} 
\maketitle

%
%

\section{Introduction}
\label{intro}
With ALICE heavy-flavour is studied in proton-proton (pp) and lead-lead (Pb-Pb) collisions at LHC energies. Due to their large masses, heavy quarks are predominantly produced via hard scatterings, in the initial phase of the collision. This makes them unique probes  since they then experience the full evolution of the strongly interacting QCD matter produced in such collisions. One of the key methods used to infer the parameters of the medium is the measurement of energy loss of the partons traversing it. One observable used to quantify the energy loss is the nuclear modification factor ($R_{\mathrm{AA}}$), which is the ratio of particle yields in heavy-ion collisions compared to pp, scaled by the number of binary nucleon-nucleon collisions. Many models predict that the heavy quarks should be less suppressed relative to their lighter flavour counterparts. The main source of the energy loss is thought to be in-medium gluon radiation, with mass and color charge dependence. The mass dependence can be found in the so called dead-cone effect, which predicts that small angle radiation is suppressed for heavy quarks \cite{deadcone}. QCD predicts less energy loss of quarks relative to gluons because of the stronger color coupling factor of the gluon \cite{colorcharge}. From these models the following order should be observed in the measured $R_{\mathrm{AA}}$: R$_{\mathrm{AA}}^{\pi}$ $<$ R$_{\mathrm{AA}}^{\mathrm{D}}$ $<$ R$_{\mathrm{AA}}^{\mathrm{B}}$. In addition to partonic energy loss, the collectivity of the QCD medium is also of interest and one observable used to quantify this is the elliptic flow ($\nu_2$), which is the azimuthal anisotropy in momentum space that arises from an "almond" shaped overlap region of a non-central heavy-ion collision expanding under anisotropic pressure gradients. For an interacting, collective medium created in a non-central collision the elliptic flow should be non-zero. For a non-interacting medium, without multiple interactions a completely isotropic distribution in momentum space would be observed. Finally, pp collisions give the baseline for Pb-Pb collisions, while also providing key tests of perturbative Quantum Chromo-Dynamic (pQCD) predictions in this new energy domain.

RHIC experiments reported the measurement of heavy quarks using heavy-flavour decay electrons from semi-leptonic decays in pp and Au-Au collisions  at $\sqrt{s_{\mathrm{NN}}}$ = 200 GeV \cite{rhice}. The measurement does not reveal the predicted dramatic differences in energy loss for heavy-flavour and at $p_{\mathrm{T}}$ greater than 4 GeV/$c$ shows a similar magnitude of suppression as light flavour in Au-Au collisions. They also reported on heavy-flavour decay electron elliptic flow and observe non-zero flow for the centrality class 20-40$\%$ \cite{phenixv2}. At LHC energies the heavy-flavour production yield is large, on the order of 50 c$\bar{\mathrm{c}}$ per event for central Pb-Pb collisions at $\sqrt{s}$ = 2.76 TeV \cite{mnr,pump,eskola}.  This allows for a more detailed study of energy loss and collectivity over a greater $p_{\mathrm{T}}$ range. The measurements are performed using multiple heavy-flavour decay channels, including the hadronic channel which allows for the first direct measurement of charm mesons in heavy-ion collisions.

%
%

\section{Heavy-flavour measurements in ALICE}
\label{hfALICE}
ALICE (A Large Ion Collider Experiment) is specifically optimized for the study of heavy-ion collisions at the LHC \cite{alice}. The performance of ALICE concerning heavy-flavour particle detection is detailed in \cite{hfper}. For the analysis presented here the detector components utilized lie in the central rapidity ($|\eta|$ $<$ 0.9), in the so-called central barrel, and at forward rapidity (-4 $<$ $\eta$ $<$ -2.5). The central barrel consists of the Inner Tracking System (ITS), surrounded by the Time Projection Chamber (TPC). Both provide high precision tracking of charged particles. Hadron identification is provided by the ionization energy loss d$E$/d$x$ in the TPC gas and information from the Time Of Flight (TOF). Electrons are also identified with TPC and TOF and at higher momentum using the ElectroMagnetic Calorimeter (EMCal) and the Transition Radiation Detector (TRD). In the forward region muon tracking and identification is provided by the muon spectrometer. In addition to the aforementioned, some small detectors are used for triggering and multiplicity measurements, including the T0 detector for time zero measurement and the V0 scintillator and Zero Degree Calorimeter (ZDC) for triggering.

In ALICE open heavy-flavour mesons are measured in the charm hadronic decay channels, where displaced vertices are selected using ITS, decay daughter particle identification is performed with TPC and TOF, and an invariant mass analysis is used. At the moment, the measured decay channels include, D$^0$ $\rightarrow$ K$^-$$\pi^+$, D$^+$ $\rightarrow$ K$^-$$\pi^+\pi^-$, D$^{*+}$ $\rightarrow$ D$^0$$\pi^+$, and D$_s^+$ $\rightarrow$ K$^+$K$^-$$\pi^+$ in $|\eta|$ $<$ 0.5. They are also measured via their semi-leptonic decay channels (D,B $\rightarrow$ $l$ + $X$) in the central region ($|\eta|$ $<$ 0.9) using TRD, EMCal, TPC, and TOF for electron identification. Here the background, dominated by Dalitz decays and photon conversions, is estimated using a Monte Carlo (MC) cocktail or e$^+$e$^-$ invariant mass method. In the forward region single muon measurement backgrounds are dominated by decays of pions and kaons. They are estimated in pp collisions using MC and are extrapolated in Pb-Pb collisions from the measured $\pi$, K yields in central rapidity.

%
%

\section{Measurements in pp collisions}
\label{pp}
The differential $p_{\mathrm{T}}$  inclusive production cross section for D$^+$ mesons in pp collisions at $\sqrt{s}$ = 7 TeV, compared with two theoretical predictions, FONLL \cite{Dfonll} and GM-VFNS \cite{gmvfns} is shown in Fig. 1. Similar measurements have been obtained for D$^0$, D$^{*+}$ \cite{xsecdmeson}, and D$_\mathrm{s}$  \cite{strange} and for D$^0$, D$^+$, and D$^{*+}$ mesons at $\sqrt{s}$ = 2.76 TeV \cite{dtwo}. The spectrum of single electrons from semi-leptonic heavy-flavour decays measured in pp collisions at $\sqrt{s}$ = 7 TeV can be seen in Fig. 2 (left) \cite{hfe}. The data are compared to FONLL theoretical predictions for inclusive semi-leptonic decays. Additionally, the spectrum of single muons from semi-leptonic heavy-flavour decays in pp collisions at $\sqrt{s}$ = 7 TeV \cite{hfm7} and at $\sqrt{s}$ = 2.76 TeV \cite{hfm} have also been measured. For the hadronic and semi-leptonic cases the results are consistent with the theoretical predictions, within the experimental and theoretical uncertainties. A similar agreement can be seen for the measured $p_{\mathrm{T}}$ differential inclusive production cross section of the three non-strange D mesons at $\sqrt{s}$ = 2.76 TeV.
\begin{figure}
\centering
\resizebox{0.50\columnwidth}{!}{%
\includegraphics{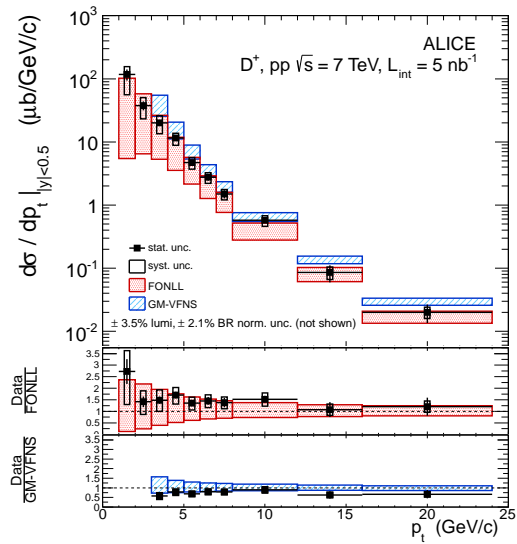} }
\caption{$p_{\mathrm{T}}$ differential cross section for D$^+$ in pp collisions at $\sqrt{s}$ = 7 TeV \cite{xsecdmeson}, compared to the theoretical predictions of FONLL \cite{Dfonll} and GM-VFNS \cite{gmvfns}.}
\label{DplusXsec}       
\end{figure}
\begin{figure}
\centering
\subfigure{
\resizebox{0.45\columnwidth}{!}{%
\includegraphics{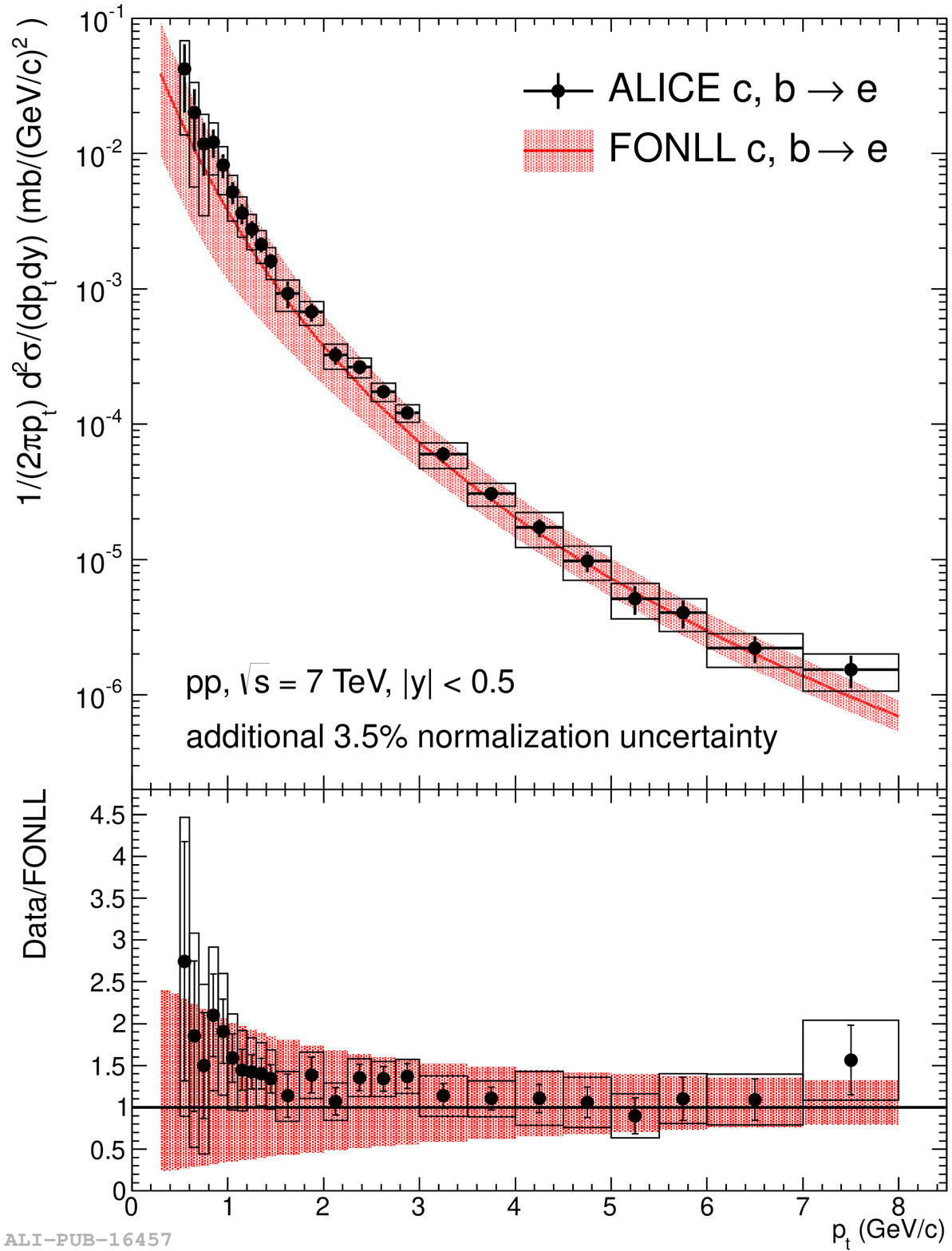} } }
\subfigure{
\resizebox{0.45\columnwidth}{!}{%
\includegraphics{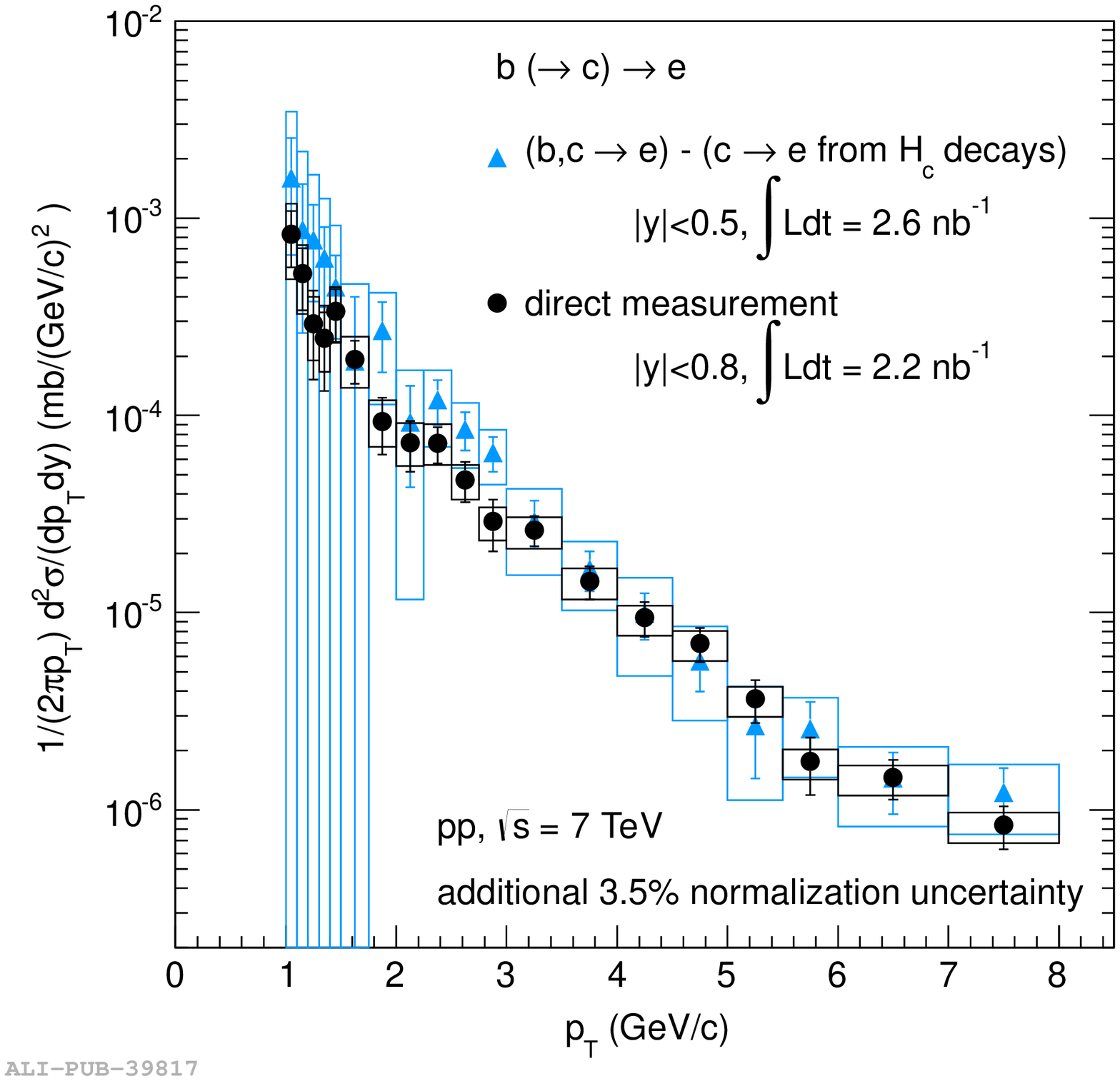} } }
\caption{Left: Electron spectrum from heavy-flavour hadron decays in pp collisions at $\sqrt{s}$ = 7 TeV (upper panel). The measurement is compared to a FONLL pQCD calculation for inclusive charm and beauty semi-leptonic decays. The ratio of the measured spectrum to the FONLL calculation is shown in the lower panel \cite{hfe}. Right: Invariant cross sections of electrons from beauty hadron decays in pp collisions at $\sqrt{s}$ = 7 TeV measured directly via the transverse impact parameter method and indirectly via subtracting the calculated charm hadron decay contribution from the measured heavy-flavour hadron decay electron spectrum \cite{mj}.}
\label{hfeXsec}       %
\end{figure}

The weighted average of the total charm production cross section was calculated using the measured D meson data at $\sqrt{s}$ = 2.76 and 7 TeV. The results agree well with measurements by the ATLAS \cite{atlas} and the LHCb collaborations \cite{lhcb}. The measurements, when compared to next-to-leading order pQCD, lie at the upper limit of the theoretical prediction. Finally, we note that the heavy-flavour single electron spectrum measured by ATLAS \cite{atlase} is consistent with the ALICE measurement.

Additional heavy-flavour measurements in pp collisions include electrons from beauty hadron decays at $\sqrt{s}$ = 7 TeV and the relative beauty contribution to the heavy-flavour electron yield at $\sqrt{s}$ = 2.76 TeV. For the former, the electrons from beauty are selected based on their displacement from the primary vertex of the collision.The production cross section of electrons from semi-leptonic decays of beauty hadrons in $|\mathrm{y}|$ $<$ 0.8 can be seen in Fig. 2 (right) \cite{mj}. The beauty contribution to the heavy-flavour electron yield is obtained using azimuthal angular correlations between heavy-flavour electrons and charged hadrons at $\sqrt{s}$ = 2.76 TeV and can be seen in Fig. 3 \cite{ehcorr}. One observes that the contribution of electrons from beauty decays is comparable to that of charm  above $p_{\mathrm{T}}$ = 5 GeV/$c$. Both measurements have been compared to FONLL pQCD calculations and agree within the uncertainties.

\begin{figure}
\centering
\resizebox{0.65\columnwidth}{!}{%
\includegraphics{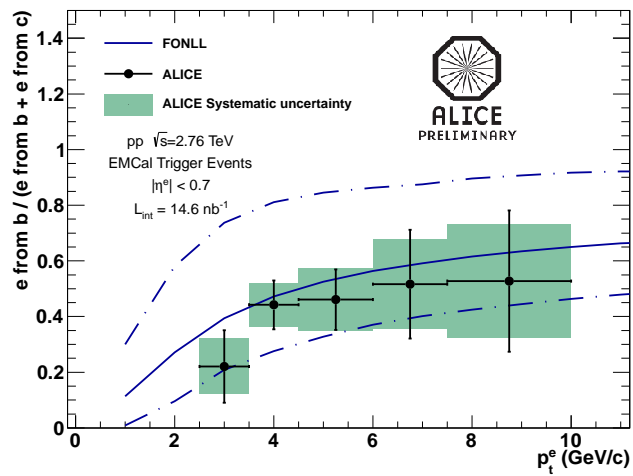} }
\caption{Relative beauty contribution to the heavy flavour electron yield in pp collisions at $\sqrt{s}$ = 2.76 TeV \cite{ehcorr} compared with next-to-leading-order pQCD calculations FONLL.}
\label{deepa}       %
\end{figure}
%

%
%

\section{Measurements in Pb-Pb collisions}
\label{PbPb}

The analysis strategy implemented for Pb-Pb data resembles that used for pp data. However, because of the high combinatorial background in the hadronic decay channel analysis, tighter selection criteria are used. For details on the published single muon and D meson analyses in 2010 Pb-Pb collisions see \cite{hfm7,DRaa2010}, respectively. The preliminary results shown in these proceedings use the 2011 Pb-Pb data.

\subsection{Nuclear modification factor $R_{\mathrm{AA}}$}

The nuclear modification factor ($R_{\mathrm{AA}}$) is defined as:

\begin{equation}
	R_{\mathrm{AA}} (p_\mathrm{T}) = \frac{1}{\langle N_{\mathrm{coll}} \rangle} \frac{\mathrm{d}N_{\mathrm{AA}}/\mathrm{d}p_\mathrm{T}}{\mathrm{d}N_{\mathrm{pp}}/\mathrm{d}p_\mathrm{T}} = \frac{1}{\langle T_{\mathrm{AA}} \rangle} \frac{\mathrm{d}N_{\mathrm{AA}}/\mathrm{d}p_\mathrm{T}}{\mathrm{d}\sigma_{\mathrm{pp}}/\mathrm{d}p_\mathrm{T}}
\end{equation}

\noindent
where d$N_{\mathrm{AA}}$/d$p_\mathrm{T}$ is the yield in Pb-Pb collisions, d$\sigma_{\mathrm{pp}}$/d$p_{\mathrm{T}}$ is the differential cross section in pp collisions, and $\langle T_{\mathrm{AA}} \rangle $ is the average nuclear overlap function and is calculated using the Glauber model assuming an inelastic nucleon-nucleon cross section of 64 mb \cite{aliceGlauber}. The measured $R_{\mathrm{AA}}$, using the 2010 Pb-Pb data set,  for the centrality classes 0-20$\%$ and 40-80$\%$ of prompt D$^0$, D$^+$, and D$^{*+}$ can be found in \cite{DRaa2010}. It shows that in central collisions (0-20$\%$) the measured suppression reaches a factor 3-4  for $p_{\mathrm{T}}$ $>$ 5 GeV/$c$. These measurements have been repeated using the 2011 Pb-Pb data set, which offers more statistics and of a higher $p_{\mathrm{T}}$ reach. Here, the $R_{\mathrm{AA}}$ is the average nuclear modification factor of all three D meson species, where the contribution of each to the average is weighted by its statistical uncertainty. Fig. 4 (left) shows the average $R_{\mathrm{AA}}$ of D mesons for the centrality class 0-7.5$\%$, in $|\mathrm{y}|$ $<$ 0.5 \cite{grelli}. First, at $p_{\mathrm{T}}$ of 7 GeV/$c$ the charmed mesons show a suppression factor of 5. Secondly, if one compares the more central 2011 D meson $R_{\mathrm{AA}}$ measurement (shown here) to the 2010 result \cite{DRaa2010} there is a hint of a larger suppression when going to more central events. Additionally shown is a comparison to the lighter species, specifically charged hadrons and pions, both measured in the centrality class 0-10$\%$. The D meson suppression is comparable and all suggest a rise at higher $p_{\mathrm{T}}$, which is consistent with the expectation that parton energy loss is weakly dependent on the parton energy, thus a decrease of the relative energy loss with increasing $p_{\mathrm{T}}$. 

Heavy-flavour decay electrons and muons have also been studied using the 2010 and 2011 Pb-Pb data sets. The muon $R_{\mathrm{AA}}$ is measured in the forward region in the range 4 $<$ $p_{\mathrm{T}}$ $<$ 10 GeV/$c$. The electrons are measured in the range 3 $<$ $p_{\mathrm{T}}$ $<$ 18 GeV/$c$ at mid-rapidity \cite{shingo}. Fig. 4 (right) shows $R_{\mathrm{AA}}$ for both the heavy-flavour decay muons and electrons in the centrality class 0-10$\%$. Both the muon and electron reach a suppression factor 3-4. It is important to note that for $p_{\mathrm{T}}$ $>$ 5 GeV/$c$ FONLL calculations predict that the leptons from B meson decays start to dominate the $p_{\mathrm{T}}$ spectrum.

\begin{figure}[ht]
\centering
\subfigure{
\resizebox{0.51\columnwidth}{!}{%
\includegraphics{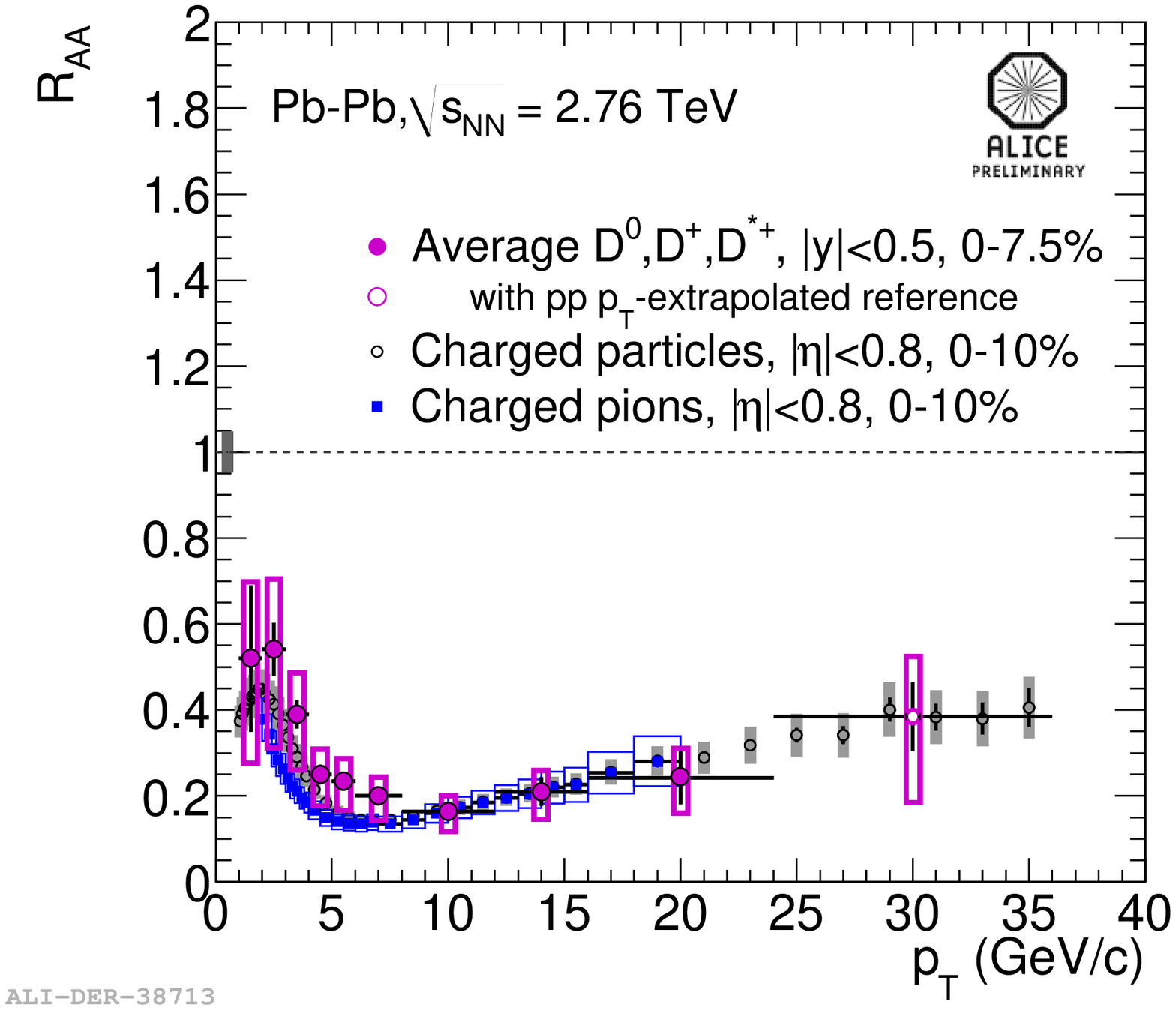}}}
\subfigure{
\resizebox{0.43\columnwidth}{!}{%
\includegraphics{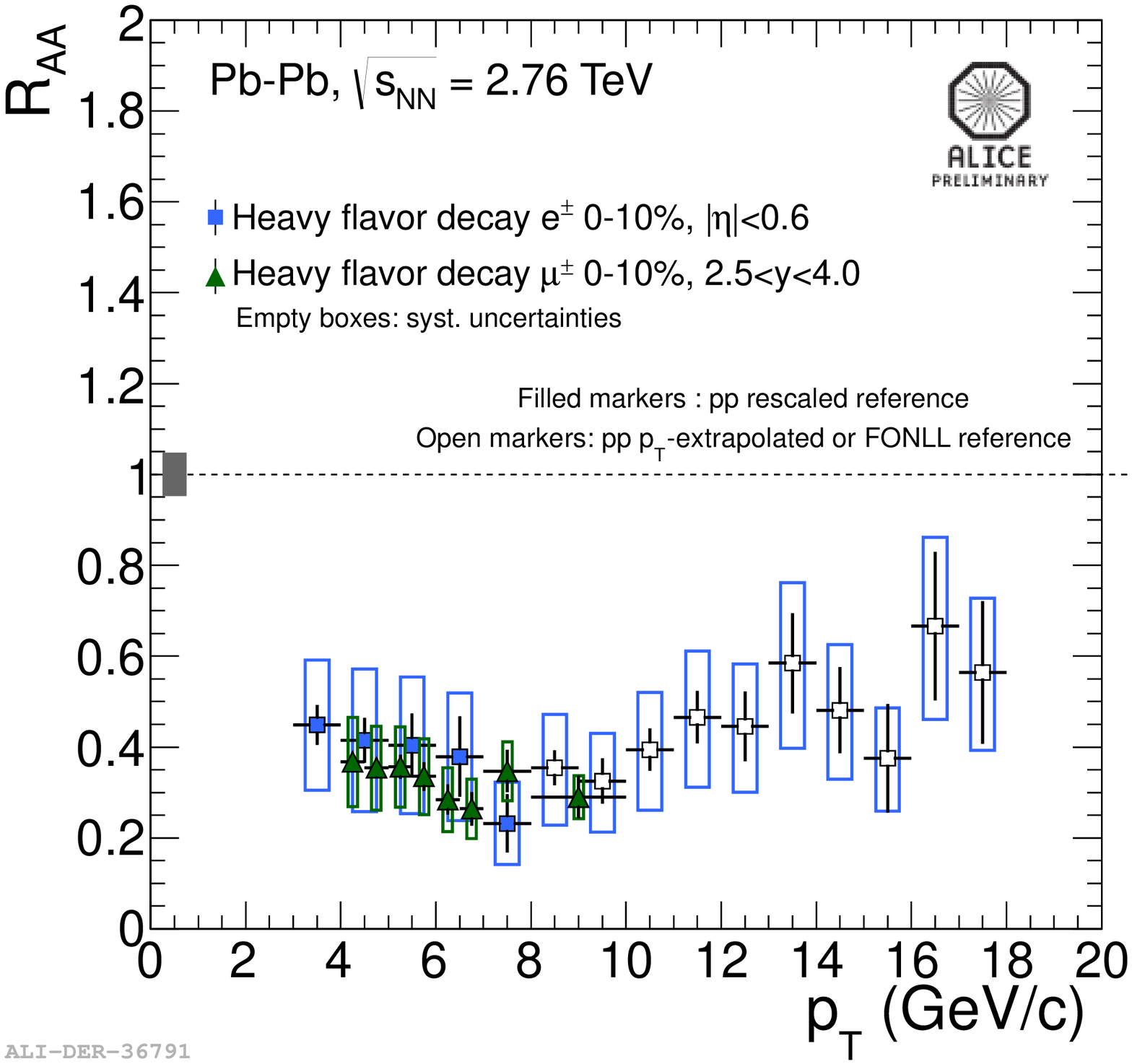} }}
\caption{Left: Average D$^0$, D$^+$ and D$^{*+}$ mesons in the 0-7.5$\%$ centrality class and charged hadron and pion $R_{\mathrm{AA}}$ in Pb-Pb collisions at $\sqrt{s_{\mathrm{NN}}}$ = 2.76 TeV in the 0-10$\%$ centrality class (2011) \cite{grelli}.  Right: Heavy flavour decay electron \cite{shingo} and muon \cite{hfm7} $R_{\mathrm{AA}}$ in Pb-Pb collisions at $\sqrt{s_{\mathrm{NN}}}$ = 2.76 TeV in the 0-10$\%$ centrality class. Vertical black lines indicate the statistical uncertainties, empty boxes the systematic uncertainties. Closed symbols represent the $R_{\mathrm{AA}}$ obtained with measured pp data (pp at 2.76TeV for muons, scaling of the 7 TeV data for electrons), while the open symbols use FONLL as pp reference. }

\end{figure}

\subsection{Elliptic flow ($\nu_2$)}

In the case of a non-central heavy-ion collision, an azimuthal anisotropy in momentum space can be observed in the measured final state particle distributions. Multiple interactions in-medium are necessary for this to occur. The particle distribution can be expanded in a Fourier series \cite{voloshin}
\begin{equation}
E \frac{d^3N}{dp^3} = \frac{1}{2\pi} \frac{d^2N}{p_T dp_T dy} \left(1 + 2 \sum_{n=1}^{\infty} \nu_n \cos [n(\phi - \Psi_{RP}] \right)
\end{equation}
\noindent
Here $\Psi_{RP}$ is the reaction plane azimuthal angle and the second coefficient, $\nu_2$, is the elliptic flow. The D meson elliptic flow was measured for the D$^0$, D$^+$, and D$^{*+}$ using the 2011 Pb-Pb data set, in the centrality class 30-50$\%$ and the preliminary result is shown in Fig. 5 (left). All three D mesons agree within statistical uncertainties and show a non-zero $\nu_2$. In this rapidity region all are comparable to the non-zero $\nu_2$ measurement of the charged hadrons (also shown in the figure) \cite{chargedv2}. The D$^0$ measurement was also performed in the centrality classes 15-30$\%$ and 0-7.5$\%$ \cite{davide}. The preliminary results are shown in Fig. 5 (right) and with increased collision centrality  the $\nu_2$ weakens, and within statistical uncertainties reaches zero in the most central case. This is consistent with the expectation that for central collisions the azimuthal momentum space anisotropy vanishes.

\begin{figure}[ht]
\centering
\subfigure{
\resizebox{0.48\columnwidth}{!}{%
\includegraphics{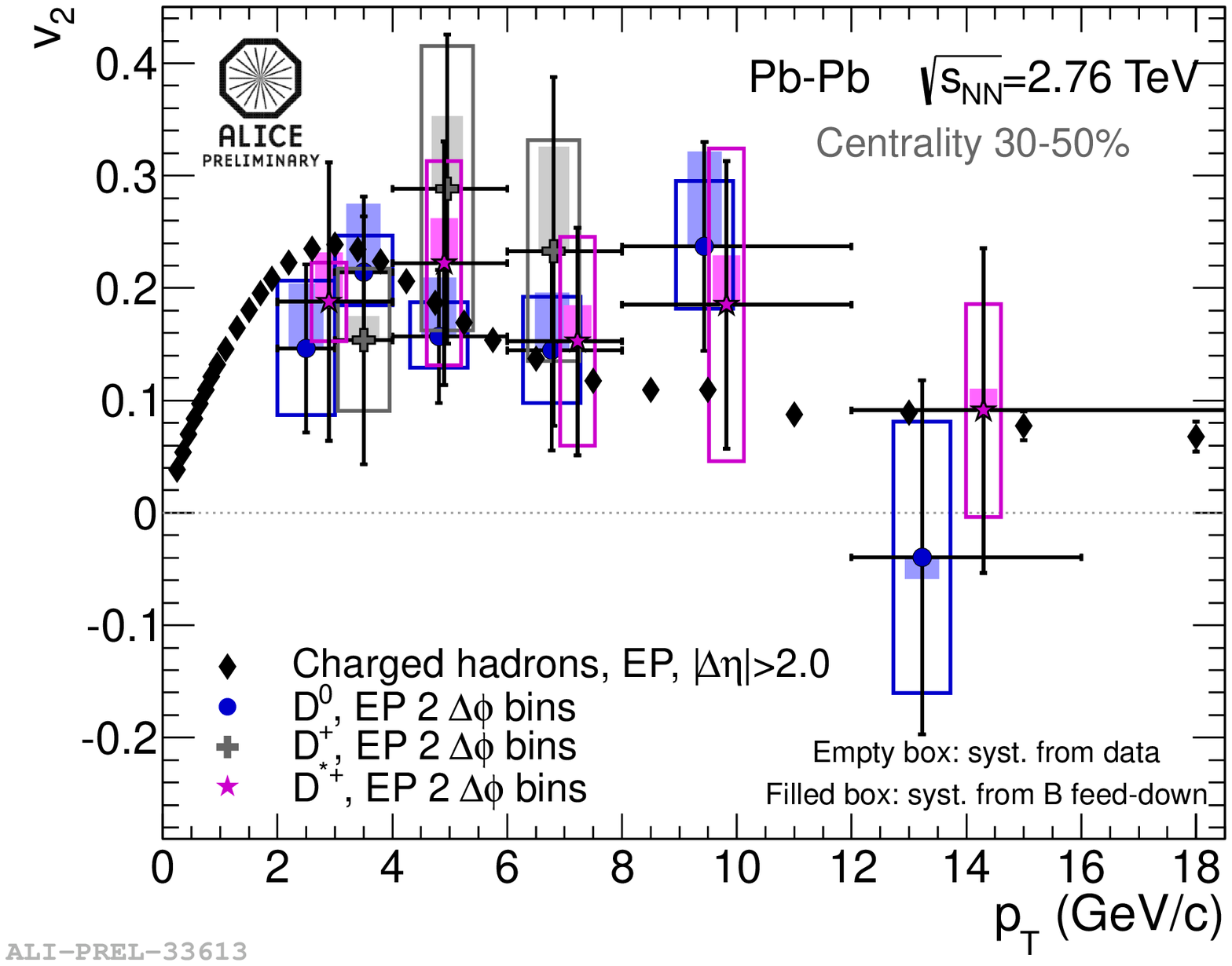} } }
\subfigure{
\resizebox{0.48\columnwidth}{!}{%
\includegraphics{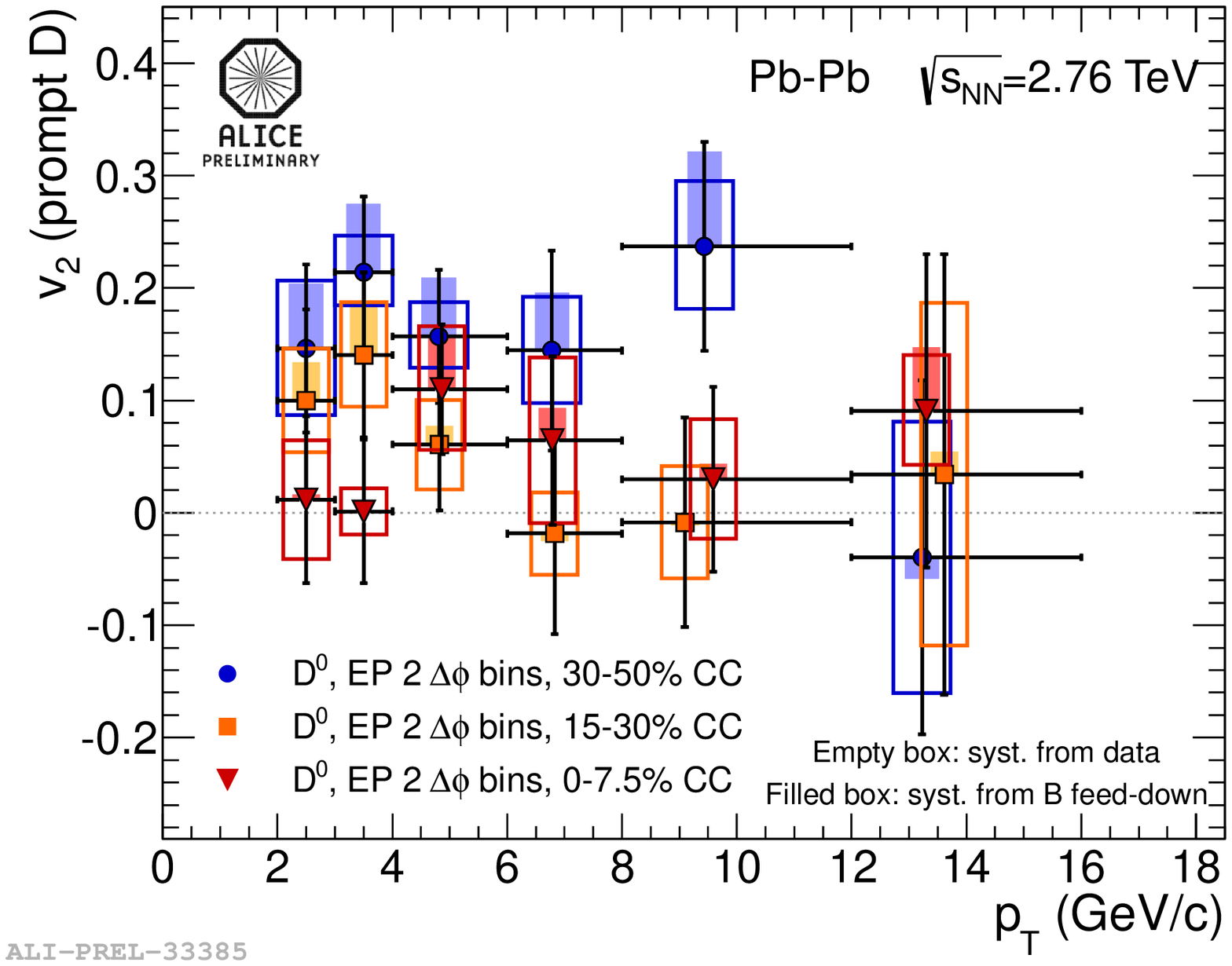} } }
\caption{Left: $\nu_2$ of D$^0$, D$^+$ and D$^{*+}$ measured in Pb-Pb collisions at $\sqrt{s_{\mathrm{NN}}}$ = 2.76 TeV in the centrality class 30-50$\%$ mcompared to charged hadron $\nu_2$. Right: $\nu_2$ of D$^0$ mesons with event plane method, superposition of 3 centrality classes: 0-7.5$\%$, 15-30$\%$, 30-50$\%$ \cite{davide}.}
\label{elliptic}
\end{figure}

\begin{figure}[ht]
\centering
\resizebox{0.48\columnwidth}{!}{%
\includegraphics{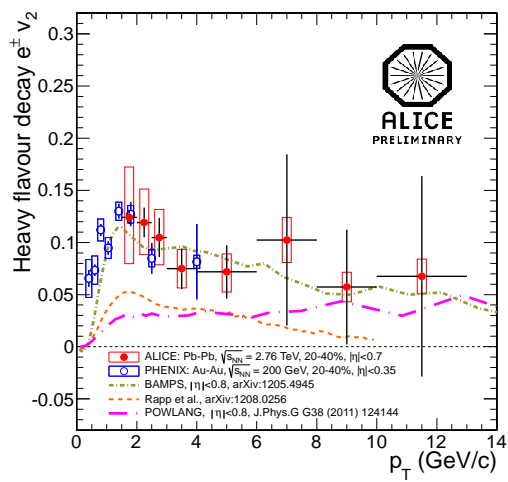} } 
\caption{Heavy-flavour electron $\nu_2$ in 20-40$\%$ Pb-Pb collisions at $\sqrt{s}$ = 2.76 TeV compared to models. See text for details.}
\label{elliptichfe}
\end{figure}

Heavy-flavour decay electron $\nu_2$ has also been reported \cite{shingo}. Fig. 6 shows the preliminary ALICE result using electrons measured in $|\eta|$ $<$ 0.7 in the centrality class 20-40$\%$, along with a comparison to the published PHENIX ($\mathrm{y}$ $<$ 0.35) $\sqrt{s_{\mathrm{NN}}}$ = 0.2 TeV Au-Au result, and model predictions. The comparison of the two different experiments at two energies shows that the magnitude of $\nu_2$, for the range 1.5 $<$ $p_{\mathrm{T}}$ $<$ 4 GeV/$c$, is consistent. The partonic transport model BAMPS \cite{bamps} prediction of heavy-flavour electron $\nu_2$ is consistent with the measurement. The heavy quark transport model with in-medium resonance scattering and coalescence (Rapp et. al.) \cite{rapp} and POWLANG,  a transport model with collisional energy loss \cite{andrea} tend to underpredict heavy-flavour electron $\nu_2$, although the sizable systematic uncertainties in the data prevent a conclusive statement at larger $p_{\mathrm{T}}$.

%
%
\subsection{Discussion}
\label{diss}

Both the D mesons, measured via their hadronic decays, and single electrons and muons, from the semi-leptonic decay of heavy-flavour hadrons, show similar behavior in the nuclear modification factor and elliptic flow to that of their lighter counterparts. In \cite{DRaa2010} ALICE compares the average D meson $R_{\mathrm{AA}}$ with several in-medium parton energy loss models that compute the nuclear modification factor. Those models which describe both the light and heavy hadrons are radiative with in-medium D meson dissociation \cite{vitev}, radiative plus collisional energy loss in the WHDG \cite{whdg}, and CUJET 1.0 \cite{cujet}. Additionally some models calculate both the nuclear modification factor and elliptic flow. Shown in Fig. 7 is the average D meson $R_{\mathrm{AA}}$ (left) and $\nu_2$ (right). A comparison of the individual models reveals the difficulty they have in describing both observables. For example, the WHDG (radiative plus collisional) implementation describes the average $R_{\mathrm{AA}}$ quite well, but significantly underestimates the $\nu_2$. While BAMPS (parton transport) is consistent with the measured $\nu_2$, but slightly over-estimates the D meson suppression.

\begin{figure}
\centering
\subfigure{
\resizebox{0.415\columnwidth}{!}{%
\includegraphics{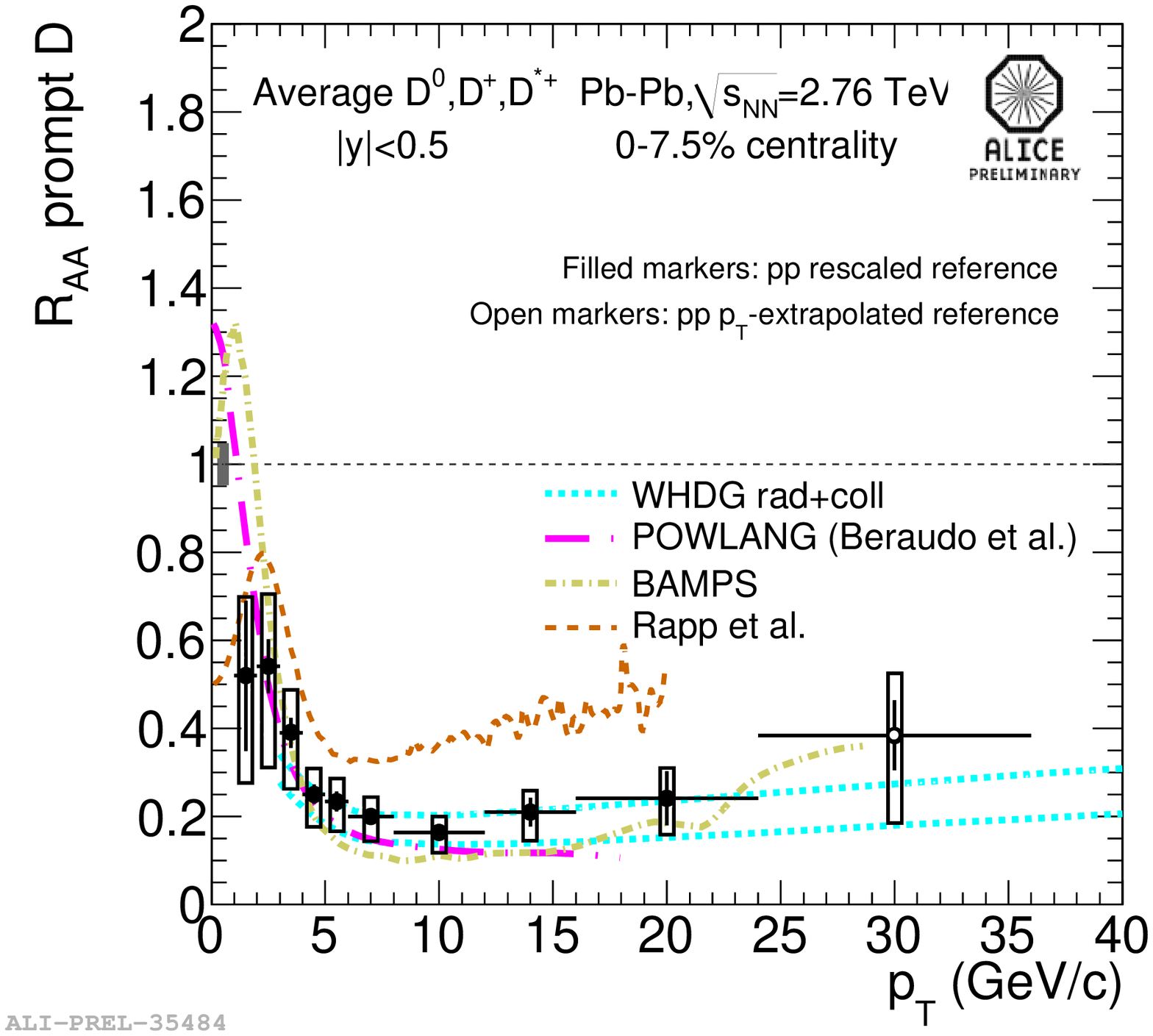}}}
\subfigure{
\resizebox{0.478\columnwidth}{!}{%
\includegraphics{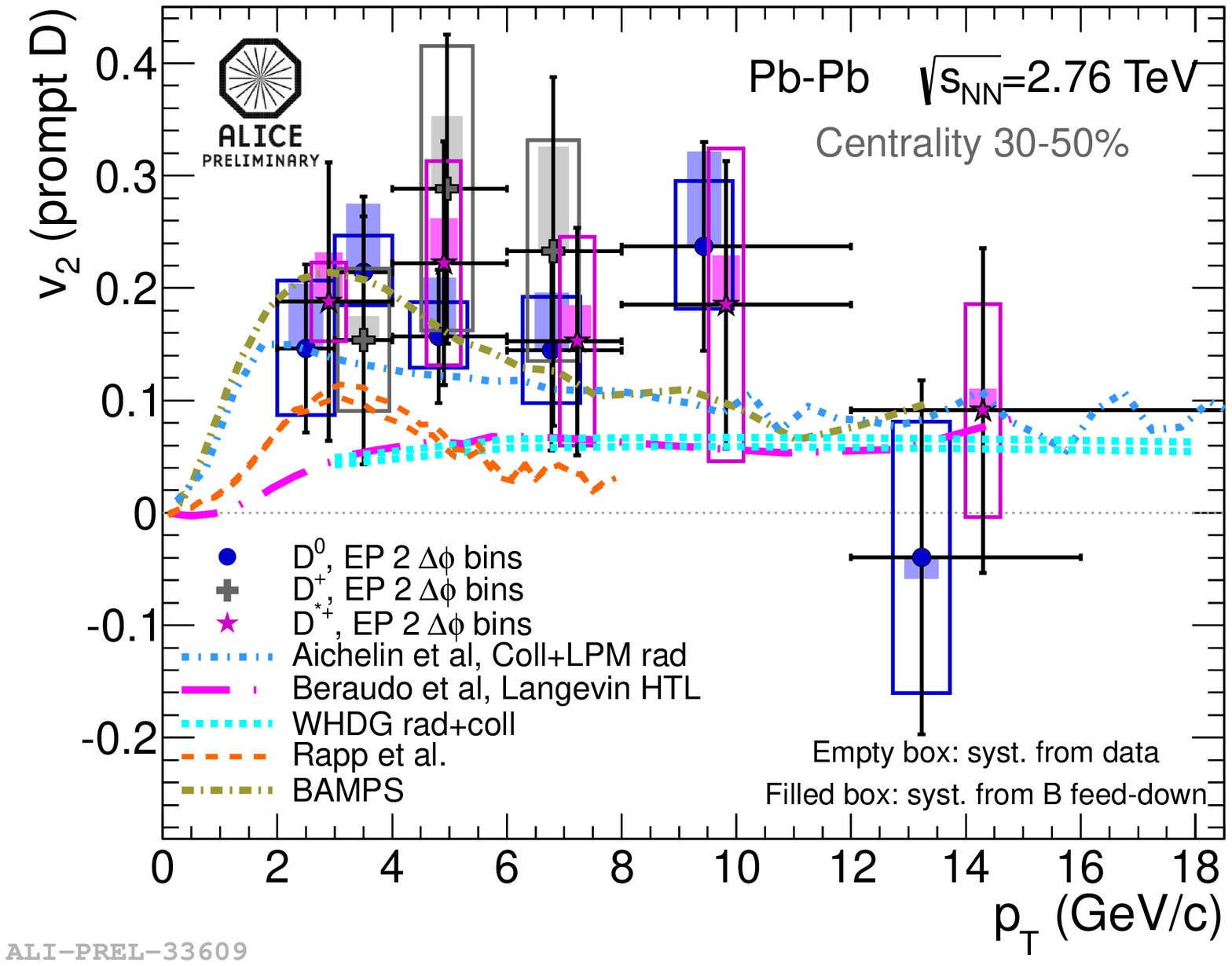}}}
\caption{Left: Average D$^0$, D$^+$ and D$^{*+}$ $R_{\mathrm{AA}}$ vs $p_{\mathrm{T}}$  in $|\mathrm{y}|$ $<$ 0.5 for 0-7.5$\%$ central Pb-Pb collisions at $\sqrt{s_{\mathrm{NN}}}$ = 2.76 TeV (2011 data). Right: $\nu_2$ of D$^0$, D$^+$ and D$^{*+}$  in centrality 30-50$\%$. Results are compared to theoretical predictions.}
\label{compare}
\end{figure}

%
%

\section{Conclusions}
\label{con}

The present status of the open heavy-flavour analyses in ALICE has been shown. For pp collisions we showed the D meson $p_{\mathrm{T}}$ differential cross section, along with the heavy flavour electron cross section. Within the uncertainties the measurements agree well with theoretical calculations. In Pb-Pb collisions the nuclear modification factor for both measurements, utilizing the hadronic and semi-leptonic decay channels, demonstrate a strong suppression, which moves towards unity in more peripheral collisions. Elliptic flow measurements of D mesons  and heavy-flavour decay electrons indicate a non-zero $\nu_2$, that is comparable to the measured charged hadrons. We also showed that in pp collisions beauty decay electrons can be measured using displaced electrons. This analysis will continue in Pb-Pb collisions with the aim of separating the charm and beauty contribution to the single electron measurement, which will give a better handle on the energy loss and level of thermalization of beauty alone. In addition, the comparison data of p-Pb collisions, scheduled for January 2013, will allow disentanglement of initial state and nuclear effects, which could differ for the light and heavy flavours.

\end{document}